# True performance metrics in beyond-intercalation batteries

Stefan A. Freunberger

**Beyond-intercalation batteries promise a step-change in energy storage compared to intercalation based lithium- and sodium-ion batteries. However, only performance metrics that include all cell components and operation parameters can tell whether a true advance over intercalation batteries has been achieved.**

Pushing energy storage beyond the limits of current batteries has become a societal demand and focus of much of the forefront research[1,2]. Lithium- and sodium-ion batteries (LIBs, SIBs) based on intercalation materials that store mobile $Li^+$ or $Na^+$ ions in stable frameworks of transition metal compounds or carbon are now approaching their limits[3]. Going beyond their limits requires lighter redox active elements that exchange more electrons and elimination of non-redox active components. This often goes along with the shift from transition to main group elements and thus better sustainability and lower cost, which together motivates potentially game-changing approaches for 'beyond-intercalation chemistries'. These approaches rely on fundamentally different reactions to intercalation to store charge. These include replacing the carbon anode of a LIB or SIB with Si or Sn alloying[4-6] or with conversion materials such as $MgH_2$ or P (ref. 7,8), or even with metallic Li or Na[2,9,10]. Intercalation cathodes may be replaced by the $O_2$ or S cathode[2,10-12]. Suitable combinations of such high-capacity electrodes may then result in higher-energy devices than LIBs.

Beyond-intercalation chemistries offer capacities of some 1000 to 4000 mAh $g^{-1}$, compared to between 100 and 350 mAh $g^{-1}$ for intercalation chemistries. Simplistic views would therefore suggest up to 10-fold increased energy in the beyond-intercalation device. These capacities are based on the mass of the formal storage material, i.e., the redox active compound that inserts/de-inserts ions such as metalated positive intercalation materials (e.g., $LiFePO_4$) or demetalated materials otherwise (e.g., carbon, Si). This calculation is often adopted in research papers since it is simple and tends to give very large numbers[2,4,10,13]. However, the formal (or active) storage material constitutes in many cases a minor and widely varying fraction of the total electrode. Numbers based on the active materials alone thus do not reflect true electrode performance and run the risk of becoming false benchmarks. Densely packed electrodes with relatively small formal capacity may, for example, have higher true capacity than loosely packed electrodes with larger formal capacity.

This Comment highlights the difficulty with reporting performances of new battery chemistries arising from the non-trivial correlation between performance with respect to the formal storage material and the full electrode. The problem is rooted in the typically much lower active material fraction in the electrode compared to intercalation materials. It is therefore useful to think of the full electrode as the ion host rather than the formal storage material. The latter is a poor indicator for true performance and the large numbers obtained make for an unjust comparison with intercalation materials. The problem is exacerbated by the large differences between battery chemistries and the often adopted practice of limiting depth-of-cycling in battery operation. To aid comparison single electrode capacities are discussed. The practically important energy and power further scale with cell voltage and the fraction of other non-active components such as separator, current collector and packaging[4,14].

**Formal host versus super-host**

Departing from the concept of LIBs/SIBs generally does not allow for stable frameworks in the active materials. Hence, high formal capacities are typically associated with large volume changes upon cycling between the demetalated and metalated phases. As shown in Fig. 1c, for example, fully alloying Si to $Li_{3.75}Si$ increases the volume by ~260% , Sn to $Na_{3.75}Sn$ by ~420%, reacting sulphur to $Li_2S$ adds 180% volume. In the $Li-O_2$ or $Na-O_2$ cathode the full volume of $Li_2O_2$ or $NaO_2$ forms/disappears during discharge/charge.

The basic charge storage processes are linking redox moieties R to electron and $Li^+$ or $Na^+$ ($M^+$) transport according to $R + x\,e^- + x\,M^+ \leftrightarrow M_xR$. However, none of the materials would allow electron and ion transport deeply into the bulk, which requires them to be used as small particles. For simultaneous contact with ionic pathways to the electrolyte and electronic pathways to the current collector, the active material particles are typically embedded into an electron conducting carbon network, held together by a binder and soaked with a liquid electrolyte. Volume stable intercalation materials can occupy a major fraction of the electrode, which is fairly similar amongst most materials and justifies comparing capacities as a first approximation based on the formal storage material mass alone. In contrast, the large volume changes of beyond-intercalation materials typically allow for much smaller and more widely varying fractions of active components. Here, inactive components can in some cases occupy several times the mass and volume of the storage material. Thus, the electrode including electrolyte becomes the actual ion host structure, which is required to link the host particles to electron and ion transport. This introduces a 'super-host structure', Fig. 1b, that becomes an indispensable and therewith integral part of the battery chemistry in a given electrode architecture and needs to be accounted for when reporting performance metrics.

Naturally, the question arises how capacities including a reasonable super-host structure compare between intercalation and beyond-intercalation chemistries. To establish a common base, it is assumed here that active material particles are spherical, which means that in the most expanded packed face centred cubic (fcc) structure, their volume occupation is 74%, the theoretically limiting value. Figure 1a and b illustrate the relationship between formal host material (red dashed domain) and super-host (black dashed box) for the example of the Si anode. Si particles are distributed in the electrode with a low enough density (Fig. 1a) to allow for expansion upon lithiation (Fig 1b). The electrolyte fills all pores around the Si particles and is displaced upon particle expansion. Figure 1d shows true capacities per mass and volume for intercalation and beyond-intercalation materials, accounting for the total mass and volume of the super host, whereas Fig. 1c only accounts for the formal material. A comparison between Fig. 1c and d shows that capacity with respect to formal material is a very poor indicator for true electrode performance and large formal capacities are bound to make an unjust impression of high capacities in comparison to intercalation materials. Beyond-intercalation offers up to 10-fold of the formal capacity per mass and volume compared to intercalation, while the gain in true capacity is only up to 3-fold for $Li^+$ and double for $Na^+$ if their volume occupation reaches 74%. A volume occupation of 74% for beyond-intercalation materials will yet be extremely difficult to achieve in practice, and the capacity in real electrodes will be even lower than those shown in Fig. 1d.

**Limited depth-of-cycling**

Owing to complexities of the storage mechanism with beyond-intercalation materials, it appears that somewhat limited depth-of-cycling – using less than the full capacity of the present storage material – may be needed to achieve the required cyclability[2,9,11,17,18]. However, care must be taken to ensure that the electrode achieves true higher capacity than an intercalation electrode. Numbers based on the mass of formal storage material tell little about true capacity. As further detailed below, if the depth-of-cycling needs to be restricted, it is vital to ensure the highest possible volume fraction of the most expanded phase.

Motivations to limit depth-of-cycling are as diverse as the chemistries. For example, the lithiation mechanism of Si is complex and depends on the voltage limits and history. Limiting capacity can be motivated to restrict the detrimental effect of volume expansion[4]. Figure 2a shows the maximum true capacity of Si alloys as a function of the formal capacity for the cases that $Li_1Si$, $Li_2Si$, or $Li_{3.75}Si$ occupy 74% of the electrode volume. It can be seen that capacity-limited cycling can still yield significant improvement over intercalation if a high volume occupation of the targeted end phase is ensured. However, if the end phase occupies a lower volume fraction, the true capacity falls quite rapidly towards the graphite benchmark. This can be seen, for example, when comparing the curve in Fig. 2a labelled $Li_1Si$ with the other curves at the same specific capacity. The reason lies in the large electrolyte/Si ratio. At 74% volume occupation of $Li_{3.75}Si$ the ratio is roughly four (Fig 2b and c); in real electrodes the ratio may be even higher. For comparison, intercalation electrodes are engineered to use significantly less electrolyte than active material as shown for graphite in Fig. 2c. Li-S cathodes face the significant challenge of very poorly conducting $S_8$ and $Li_2S$, which means that the full utilization of the available sulphur is problematic even if it is embedded in a large amount of carbon and electrolyte. Poor sulphur utilization together with a large inactive material to sulphur ratio curbs true energy quickly below LIBs[2,12,19,20]. Cycling Li and Na metal is still associated with relatively poor coulombic efficiencies despite considerable progress recently[2]. Metals and electrolytes are consumed due to the irreversibility with metal plating/stripping and they need to be over-supplied, but then the ratio of inactive material to actively cycled metal becomes large.

Metal-$O_2$ batteries are special in that the cathode in the charged state does not contain the redox material that could be taken as a reference for capacity. Thus, it is convenient to report capacities per weight of porous electrode. The capacity at a given initial porosity is determined by the degree of pore filling with $Li_2O_2$ or $NaO_2$. The large numbers that are often reached as first discharge capacities (up to several 10,000 mAh $g_{carbon}^{-1}$) compare superficially favourably with intercalation materials (some 100 mAh $g^{-1}$). At the fcc packing density the capacity is ~700 mAh $g_{total}^{-1}$, which is still much higher than intercalation electrodes, Fig. 1d and Fig. 2d-f. As a result of the problems to cycle at full capacity it has become a habit to report cycling at, e.g., 1,000 mAh $g_{carbon}^{-1}$, which may still seem a lot in comparison to LIBs[17,18]. Seemingly equal formal capacities are, however, easily misjudged since true capacities strongly depend on initial porosity and thus the substrate/electrolyte ratio; the black squares and red circles in Fig. 2d show increasing true capacity with decreasing initial porosity (by varying carbon content) at 1,000 mAh $g_{carbon}^{-1}$. This is because of the large electrolyte/$Li_2O_2$ ratio at shallow discharge (Fig. 2e and f). In some cases this limited capacity cycling allows simulation of a large possible cycle number despite the cumulative capacity of these cycles hardly exceeding the capacity of a few full cycles.

Clearly, overly limited cycling regimes are unsuitable to demonstrate large reversible capacities for many cycles in all cell chemistries. Yet, it is a common feature of beyond-intercalation chemistries that reasonably capacity-limited cycling can enable cyclability and at the same time yield significant improvement over intercalation if the capacity per total weight is kept in mind as shown in Fig 2a and d.

**Reporting true electrode performance**

When performance is the argument for research work then data need to support it. To achieve higher true capacities than intercalation electrodes it is crucial with all beyond-intercalation materials to achieve an as high as possible packing density, and to maximize the utilization of active material but minimize that of inactive materials. Low packing density and restricted depth-of-cycling likely result in no advantage over intercalation electrodes as demonstrated in Fig. 2. Reporting capacity with respect to formal material mass does not reveal whether the electrode performs better than an intercalation electrode. A fair assessment therefore requires giving capacity per total electrode mass and volume.

Unfortunately, many studies do not report the required measures to work out full electrode performance metrics. Figure 1 and 2 demonstrate that these measures are the porous electrode volume as well as the mass and volume fractions of carbon, binder, electrolyte, and active material. As pointed out in Box 1, they are easily obtained and it is recommended that they are reported in research papers. The following three points should be further emphasised. First, capacity per active material mass alone is inherently much larger for beyond-intercalation than for intercalation materials; it only gives a measure of active material utilization. Second, the correlation between capacity with respect to active material and total electrode is not obvious and varies vastly amongst cell chemistries and electrode architectures. Third, only beyond-intercalation electrodes with higher true capacity than intercalation electrodes will raise device performance.

There is no theoretical barrier for beyond-intercalation chemistries to achieve higher energies than LIBs/SIBs in practice. However, it is important to present performance data in a consistent manner with respect to full electrode mass and volume so as to facilitate and appropriately track progress in the field. This is essential to allow for a fair assessment of cell performance including energy, power and cycle life.

*Stefan A. Freunberger is at Graz University of Technology, Stremayrgasse 9, 8010 Graz, Austria. e-mail: freunberger@tugraz.at*

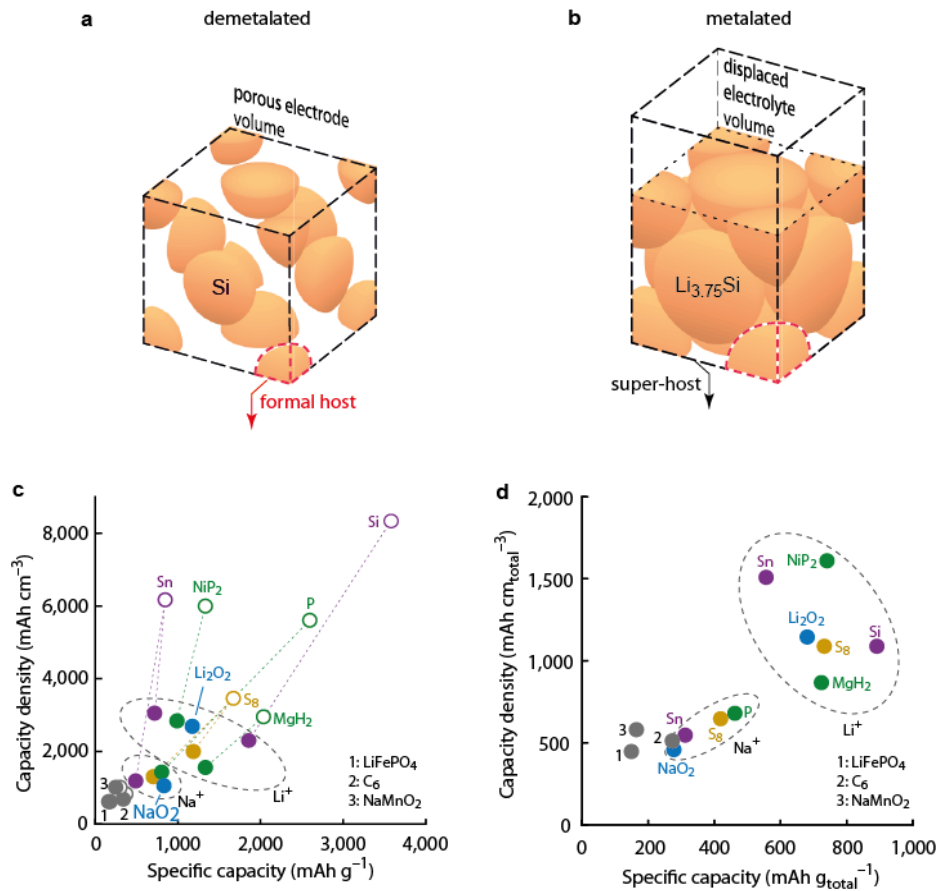

**Figure 1 | Capacities with respect to the formal storage material and the super-host structure**. **a**, **b**, Unit cell of the super-host structure for face centred cubic (fcc) close packed spheres of the formal storage material in the demetalated state (**a**) and metalated state (**b**) for the example of Si/Li$_{3.75}$Si. In the demetalated state the void space is filled by electrolyte, which is displaced upon expansion of the storage material. Sizes of spheres are to scale. The balance domains for formal and true capacity are denoted by the red dashed domain for formal material and the black dashed box for the super-host, respectively. **c**, Formal capacity per weight and volume of metalated (filled symbols) and demetalated (open symbols) phases of a selection of storage materials. Colours denote intercalation materials (grey), alloying (violet), sulphides (yellow), oxides (blue), and conversion materials (green). Capacities were calculated for the couples FePO$_4$/LiFePO$_4$, C$_6$/LiC$_6$, β-Na$_0$MnO$_2$/β-NaMnO$_2$ (Ref. 15) Si/Li$_{3.75}$Si, Sn/Li$_{3.75}$Sn, Sn/Na$_{3.75}$Sn, S$_8$/Li$_2$S, S$_8$/Na$_2$S, O$_2$/Li$_2$O$_2$, O$_2$/NaO$_2$, MgH$_2$/Mg + 2 LiH, NiP$_2$/Ni + 2 Li$_3$P, P/Na$_3$P from their density and formula weight, and normalized to the masses and volumes of the demetalated/metalated phases. Densities were obtained from the crystal structures, or published values for the alloys[5,16]. **d**, Capacities per weight and volume for the same materials including super-host structure and electrolyte. Carbon and binder account for 4% volume each. Values were normalized to the metalated state with the electrolyte volume filling the void in the demetalated state. The dashed regions are guides to the eye to indicate the areas for Li$^+$ and Na$^+$ storage materials.

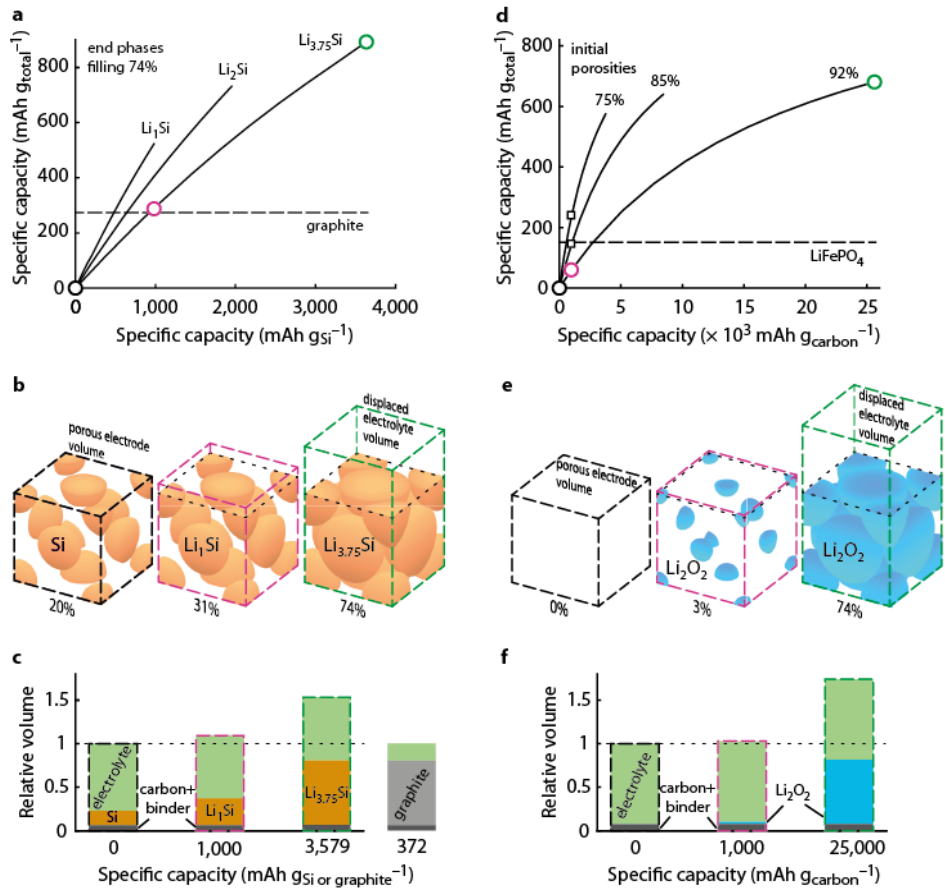

**Figure 2 | True electrode capacity with limited capacity cycling**. **a**, True capacity of a Si alloying anode as a function of Si utilization when either $Li_1Si$, $Li_2Si$, or $Li_{3.75}Si$ fills 74% of the electrode volume (given by the labels above the curves). The analogous value for graphite as standard intercalation anode is given by the dashed line for comparison. Carbon and binder account for 4% volume each. **b**, Space filling of spherical active material particles inside the porous electrode volume and the displaced electrolyte volume (together with the super-host structure) for Si, $Li_1Si$, and $Li_{3.75}Si$ (0, ~1,000, and 3,579 mAh $g_{Si}^{-1}$, respectively; indicated by the three circles in **a**). Sphere sizes are to scale and the percentages indicate their volume occupation in the porous electrode. **c**, Volumes of the electrode components at these lithiation states normalized to the delithiated state. Values for a graphite anode are shown to demonstrate a very different electrolyte/active material ratio. **d**, True capacity of a $Li$-$O_2$ cathode as a function of capacity per mass of carbon for three cases of initial porosity (given by percentages above the curves). The binder accounts for 4% of the volume, and the carbon volume fraction is adapted to yield the initial porosity. At an initial porosity of 92%, fcc packing of $Li_2O_2$ corresponds to 80% pore filling. The same 80% pore filling is assumed for the other initial porosities. The black squares and red circles at 1,000 mAh $g_{carbon}^{-1}$ illustrate that respective true capacities vary strongly with electrode architecture. The analogous value for the intercalation material $LiFePO_4$ is given by the dashed line for comparison. **e**, Space filling of spherical $Li_2O_2$ particles inside the porous electrode and the displaced electrolyte volume (together with the super-host structure) at 0, 1,000, and 25,000 mAh $g_{carbon}^{-1}$, respectively (indicated by the circles in **d**). **f**, Volumes of the electrode components at these capacities normalized to the full electrode volume in the delithiated state.

**Box 1 | Three recommendations for reporting experimental parameters.**

1. The thickness of the porous electrode with the formal charge storage material in the demetalated and metalated state.
2. The mass fractions of all electrode components (carbon, binder, active material, and electrolyte), which can be obtained from the known mass fractions of all solids, their loading and the loading of electrolyte per unit electrode area.
3. The volume fractions of all electrode components, which can be obtained from the mass fractions and the respective densities.

The key is high active material packing density and a small inactive/active material ratio. With these measures it is straightforward to convert the capacity with respect to active material mass into true capacity per mass and volume of the total electrode including electrolyte in the expanded (metalated) state. Theoretically limiting cases obtained along this line for selected beyond-intercalation materials are discussed in Fig. 1b and d, and Fig. 2.